\documentclass[aps,prl,twocolumn,superscriptaddress,showpacs]{revtex4}

\usepackage{graphicx}
\usepackage{dcolumn}
\usepackage{bm}
\usepackage{amsmath}
\usepackage{amssymb}
\usepackage{lineno}
\usepackage{setspace}

\newcommand{\ppbar}{\ensuremath{p\bar{p}\,}}
\newcommand{\met}{$E_{\rm{T}}\hspace{-1.1em}/$\hspace{0.7em}}

\def \pp  {$p\bar{p}$~}
\def \tt  {$t\bar{t}$~}
\def \Mtt {$M_{t\bar{t}}$~}

\begin{document}
%
%
\title{Search for Resonant \tt Production  in \ppbar ~Collisions at $\sqrt {s} =
1.96$~TeV}
%
\affiliation{Institute of Physics, Academia Sinica, Taipei, Taiwan 11529, Republic of China} 
\affiliation{Argonne National Laboratory, Argonne, Illinois 60439} 
\affiliation{Institut de Fisica d'Altes Energies, Universitat Autonoma de Barcelona, E-08193, Bellaterra (Barcelona), Spain} 
\affiliation{Baylor University, Waco, Texas  76798} 
\affiliation{Istituto Nazionale di Fisica Nucleare, University of Bologna, I-40127 Bologna, Italy} 
\affiliation{Brandeis University, Waltham, Massachusetts 02254} 
\affiliation{University of California, Davis, Davis, California  95616} 
\affiliation{University of California, Los Angeles, Los Angeles, California  90024} 
\affiliation{University of California, San Diego, La Jolla, California  92093} 
\affiliation{University of California, Santa Barbara, Santa Barbara, California 93106} 
\affiliation{Instituto de Fisica de Cantabria, CSIC-University of Cantabria, 39005 Santander, Spain} 
\affiliation{Carnegie Mellon University, Pittsburgh, PA  15213} 
\affiliation{Enrico Fermi Institute, University of Chicago, Chicago, Illinois 60637} 
\affiliation{Comenius University, 842 48 Bratislava, Slovakia; Institute of Experimental Physics, 040 01 Kosice, Slovakia} 
\affiliation{Joint Institute for Nuclear Research, RU-141980 Dubna, Russia} 
\affiliation{Duke University, Durham, North Carolina  27708} 
\affiliation{Fermi National Accelerator Laboratory, Batavia, Illinois 60510} 
\affiliation{University of Florida, Gainesville, Florida  32611} 
\affiliation{Laboratori Nazionali di Frascati, Istituto Nazionale di Fisica Nucleare, I-00044 Frascati, Italy} 
\affiliation{University of Geneva, CH-1211 Geneva 4, Switzerland} 
\affiliation{Glasgow University, Glasgow G12 8QQ, United Kingdom} 
\affiliation{Harvard University, Cambridge, Massachusetts 02138} 
\affiliation{Division of High Energy Physics, Department of Physics, University of Helsinki and Helsinki Institute of Physics, FIN-00014, Helsinki, Finland} 
\affiliation{University of Illinois, Urbana, Illinois 61801} 
\affiliation{The Johns Hopkins University, Baltimore, Maryland 21218} 
\affiliation{Institut f\"{u}r Experimentelle Kernphysik, Universit\"{a}t Karlsruhe, 76128 Karlsruhe, Germany} 
\affiliation{Center for High Energy Physics: Kyungpook National University, Taegu 702-701, Korea; Seoul National University, Seoul 151-742, Korea; SungKyunKwan University, Suwon 440-746, Korea; Korea Institute of Science and Technology Information, Daejeon, 305-806, Korea; Chonnam National University, Gwangju, 500-757, Korea} 
\affiliation{Ernest Orlando Lawrence Berkeley National Laboratory, Berkeley, California 94720} 
\affiliation{University of Liverpool, Liverpool L69 7ZE, United Kingdom} 
\affiliation{University College London, London WC1E 6BT, United Kingdom} 
\affiliation{Centro de Investigaciones Energeticas Medioambientales y Tecnologicas, E-28040 Madrid, Spain} 
\affiliation{Massachusetts Institute of Technology, Cambridge, Massachusetts  02139} 
\affiliation{Institute of Particle Physics: McGill University, Montr\'{e}al, Canada H3A~2T8; and University of Toronto, Toronto, Canada M5S~1A7} 
\affiliation{University of Michigan, Ann Arbor, Michigan 48109} 
\affiliation{Michigan State University, East Lansing, Michigan  48824} 
\affiliation{University of New Mexico, Albuquerque, New Mexico 87131} 
\affiliation{Northwestern University, Evanston, Illinois  60208} 
\affiliation{The Ohio State University, Columbus, Ohio  43210} 
\affiliation{Okayama University, Okayama 700-8530, Japan} 
\affiliation{Osaka City University, Osaka 588, Japan} 
\affiliation{University of Oxford, Oxford OX1 3RH, United Kingdom} 
\affiliation{University of Padova, Istituto Nazionale di Fisica Nucleare, Sezione di Padova-Trento, I-35131 Padova, Italy} 
\affiliation{LPNHE, Universite Pierre et Marie Curie/IN2P3-CNRS, UMR7585, Paris, F-75252 France} 
\affiliation{University of Pennsylvania, Philadelphia, Pennsylvania 19104} 
\affiliation{Istituto Nazionale di Fisica Nucleare Pisa, Universities of Pisa, Siena and Scuola Normale Superiore, I-56127 Pisa, Italy} 
\affiliation{University of Pittsburgh, Pittsburgh, Pennsylvania 15260} 
\affiliation{Purdue University, West Lafayette, Indiana 47907} 
\affiliation{University of Rochester, Rochester, New York 14627} 
\affiliation{The Rockefeller University, New York, New York 10021} 
\affiliation{Istituto Nazionale di Fisica Nucleare, Sezione di Roma 1, University of Rome ``La Sapienza," I-00185 Roma, Italy} 
\affiliation{Rutgers University, Piscataway, New Jersey 08855} 
\affiliation{Texas A\&M University, College Station, Texas 77843} 
\affiliation{Istituto Nazionale di Fisica Nucleare, University of Trieste/\ Udine, Italy} 
\affiliation{University of Tsukuba, Tsukuba, Ibaraki 305, Japan} 
\affiliation{Tufts University, Medford, Massachusetts 02155} 
\affiliation{Waseda University, Tokyo 169, Japan} 
\affiliation{Wayne State University, Detroit, Michigan  48201} 
\affiliation{University of Wisconsin, Madison, Wisconsin 53706} 
\affiliation{Yale University, New Haven, Connecticut 06520} 
\author{T.~Aaltonen}
\affiliation{Division of High Energy Physics, Department of Physics, University of Helsinki and Helsinki Institute of Physics, FIN-00014, Helsinki, Finland}
\author{A.~Abulencia}
\affiliation{University of Illinois, Urbana, Illinois 61801}
\author{J.~Adelman}
\affiliation{Enrico Fermi Institute, University of Chicago, Chicago, Illinois 60637}
\author{T.~Akimoto}
\affiliation{University of Tsukuba, Tsukuba, Ibaraki 305, Japan}
\author{M.G.~Albrow}
\affiliation{Fermi National Accelerator Laboratory, Batavia, Illinois 60510}
\author{B.~\'{A}lvarez~Gonz\'{a}lez}
\affiliation{Instituto de Fisica de Cantabria, CSIC-University of Cantabria, 39005 Santander, Spain}
\author{S.~Amerio}
\affiliation{University of Padova, Istituto Nazionale di Fisica Nucleare, Sezione di Padova-Trento, I-35131 Padova, Italy}
\author{D.~Amidei}
\affiliation{University of Michigan, Ann Arbor, Michigan 48109}
\author{A.~Anastassov}
\affiliation{Rutgers University, Piscataway, New Jersey 08855}
\author{A.~Annovi}
\affiliation{Laboratori Nazionali di Frascati, Istituto Nazionale di Fisica Nucleare, I-00044 Frascati, Italy}
% ...
\collaboration{CDF Collaboration\footnote{With visitors from $^a$University of Athens, 15784 Athens, Greece, 
$^b$Chinese Academy of Sciences, Beijing 100864, China, 
$^c$University of Bristol, Bristol BS8 1TL, United Kingdom, 
$^d$University Libre de Bruxelles, B-1050 Brussels, Belgium, 
$^e$University of California, Irvine, Irvine, CA  92697, 
$^f$University of California Santa Cruz, Santa Cruz, CA  95064, 
$^g$Cornell University, Ithaca, NY  14853, 
$^h$University of Cyprus, Nicosia CY-1678, Cyprus, 
$^i$University College Dublin, Dublin 4, Ireland, 
$^j$University of Edinburgh, Edinburgh EH9 3JZ, United Kingdom, 
$^k$University of Heidelberg, D-69120 Heidelberg, Germany, 
$^l$Universidad Iberoamericana, Mexico D.F., Mexico, 
$^m$University of Manchester, Manchester M13 9PL, England, 
$^n$Nagasaki Institute of Applied Science, Nagasaki, Japan, 
$^o$University de Oviedo, E-33007 Oviedo, Spain, 
$^p$Queen Mary's College, University of London, London, E1 4NS, England, 
$^q$Texas Tech University, Lubbock, TX  79409, 
$^r$IFIC(CSIC-Universitat de Valencia), 46071 Valencia, Spain, 
}}
\noaffiliation

%
%
\date{\today}
%
\begin{abstract}
%
%
We report on a search for narrow-width particles decaying 
to a top and antitop quark pair. The data set used in the analysis 
corresponds to an integrated luminosity of $680 ~\rm{pb^{-1}}$ collected 
with the Collider Detector at Fermilab in Run II. We present 95\% 
confidence level upper limits on the cross section times branching ratio.
Assuming a specific topcolor-assisted technicolor production model, 
the leptophobic $Z'$ with width $\Gamma_{Z'}=0.012M_{Z'}$, we exclude 
the mass range $M_{Z'} < 725~{\rm{GeV}}/c^2$ at the 95\% confidence level.
\end{abstract}
%
%
\maketitle
%
%
The discovery of the top quark in 1995 at Fermilab \cite{TopDiscovery} 
concluded a long search which occupied particle physicists for about two 
decades. Nevertheless, a decade after the top quark discovery there is 
still much to investigate about this massive but fundamental particle.
There are several ways in which new 
physics can make an appearance through the top quark.
The large mass in particular suggests that the top quark may play a 
special role in the dynamics of electroweak symmetry breaking and act 
as a powerful probe in this physics sector. 

Among the technicolor scenarios, \cite{Eichten} predicts
the existence of a heavy meson $\eta_T$ produced in 
gluon-gluon interactions
($gg \to \eta_T \to t\bar t$).
Topcolor models were investigated even before the actual discovery 
of the top \cite{Hill1}, and then extended to topcolor assisted
technicolor \cite{Hill2}.
These scenarios account for
the spontaneous breaking of electroweak symmetry by introducing 
new strong dynamics, which would explain the large top quark mass.
An SU(3) $\times$ SU(1) residual global symmetry below 1 TeV
predicts the existence of new particles like ``topgluons'' and
a color singlet vector particle ($Z'$) which couples primarily
to the third quark generation. The existence of 
a narrow leptophobic $Z'$ 
decaying to \tt has been predicted in \cite{Harris}
and searched for in Tevatron Run I data \cite{Run1X0}.
%

Other \tt production mechanisms can be found in universal extra dimensions
theories.
For example \cite{Dobrescu} investigated a standard model in 5+1 dimensions,
where 
a series of narrow resonances 
decay to \tt final states.

In this Letter, we present a search for narrow
resonances decaying to \tt by looking for an anomalous peak
in the 
\tt invariant mass distribution.
%
The search algorithm was developed before looking at
the data and optimized for signal from a leptophobic $Z'$ 
decay. 
At the Tevatron, the $Z'$ resonance can be produced via $q\bar q$ annihilation.
The standard model (SM) predicts that the top quark decays more than 99\%
of the time into a $Wb$ pair and then the $W$ can decay either leptonically
($BR = 32.4\%$)
or hadronically ($BR = 67.6\%$) \cite{PDG}.

The Collider Detector at Fermilab II (CDFII) \cite{cdf} is an azimuthally and forward-backward
symmetric apparatus designed to study \pp collisions at $\sqrt{s} =
1.96$~TeV at the Tevatron.
The detector has a charged particle tracking system \cite{Tracking}
in a 1.4 T  magnetic field directed parallel to the
proton beam direction. This system consists of a silicon microstrip detector, which 
covers
the radial range from 1.35 cm to 28 cm, and an
open-cell drift chamber
in the radial range from 40 cm to 137 cm. 
Outside the tracking system are the
electromagnetic and hadronic calorimeters, 
with projective segmentation in $\eta$-$\phi$  \cite{XYZ}.
A set of drift chambers and scintillation counters detects muons 
in the central region ($|\eta|<1$). 
The beam luminosity is determined by measuring the inelastic \pp
collision rate with gas Cherenkov detectors \cite{CLC} located in the 
$3.7<|\eta|<4.7$ region.
%
%

We search for events in which
one $W$ boson decays hadronically and one leptonically. 
The partonic final state for this decay channel is
$\ell \nu_{\ell} b \bar{b} q \bar{q}'$ ($\ell$=e, $\mu$), therefore
the final state events have one high momentum charged lepton,
large missing transverse energy (\met) 
\cite{XYZ} 
due to the 
undetected neutrino, and 
four jets from the quarks' parton shower and hadronization processes
(``lepton+jets'' channel).

The data sample for this analysis has been collected using
high-$p_T$ triggers for electrons and muons \cite{cdf}. The offline
selection requires an electron (or muon) candidate with 
$E_{\rm T} (p_{\rm T})> 20 ~{\rm{GeV}}$
contained in the central detector region.
A jet is defined as an energy cluster in the
calorimeter and is reconstructed using a fixed 
cone algorithm with a cone of radius
0.4 in $\eta-\phi$ space.
Following the procedure described in \cite{jetcorr}, energies and
\met\, are corrected to account for inhomogeneities in the
detector response and multiple \pp collisions in an event. After these
corrections we select events containing at least 4 jets with $|\eta|<2$ and 
$E_{\rm T} > 15 ~\rm{GeV}$, and with \met $> 20 ~\rm{GeV}$. 
Jets originating from the hadronization of $b$ quarks
are identified (\emph{b}-tagged), with about 40\% efficiency, 
by reconstructing their displaced vertices \cite{SecVtx}.
The tagging information is not used for event selection, 
but only to reduce jet combinatorics during event reconstruction.

Standard model processes that result in the same final state
as the resonance are backgrounds to this search. The dominant
components are: $W$ boson production in association
with jets (\emph{W}+jets), \tt production, multi-jet
events where a jet is mis-identified as a charged lepton 
and missing energy is generated by jet energy mismeasurement,
and diboson production ($WW$, $WZ$, $ZZ$) with extra jets from initial 
and final state radiation (ISR, FSR). The expected contributions in a data set 
corresponding to an integrated luminosity of $680~\rm{pb^{-1}}$
are presented in Table \ref{tab:Backgrounds}.
\begin{table}[htbp]
\caption{Expected number of events from standard model processes in a dataset
corresponding to an integrated luminosity of $680 ~ \rm{pb^{-1}}$ of data,
assuming the null hypothesis. The observed number of events on data is 450.}
\begin{center}
\begin{tabular}{c c}
\hline
\hline
Background     & number of events  \\
\hline
SM~ $t\bar t$  			&   199 $\pm$ 24 \\
diboson      			&   14  $\pm$ 1	 \\
multi-jet          		&   22  $\pm$ 10 \\
W$(\to \ell\bar\nu_\ell)+4j $ 	&   215 $\pm$ 30 \\
\hline
\hline
\end{tabular}
\label{tab:Backgrounds}
\end{center}
\end{table}
Standard model \tt and diboson expectations are calculated
according to their
theoretical cross sections \cite{ttXsec,WWXsec} and their
acceptances estimated from Monte Carlo simulations (see later). 
The quoted uncertainties
include the uncertainties on the theoretical cross sections, 
integrated luminosity, and acceptances.
Given the lack of precise theoretical predictions for 
\emph{W}+jets and multi-jet 
production cross sections, the contributions from these backgrounds
listed in Table \ref{tab:Backgrounds} are obtained by constraining the
total numbers to add up to the 450 events observed in data.
The multi-jets to \emph{W}+jets ratio 
in this sample is fixed at 10\%, which
is consistent with an estimate obtained in a previous
CDF analysis \cite{cdf}. 
The quoted uncertainties for these two backgrounds 
account for both the Poisson statistical uncertainty on the
total number of observed events and the uncertainties
on the \tt and diboson predictions.
%
%

The signature for top-quark-pair resonant production
is a peak in the invariant mass spectrum of the \tt pair ($M_{t\bar{t}}$),
over a smooth background from the standard model contributions.
In the mass range investigated here, all of the
standard model backgrounds fall smoothly with increasing
$M_{t\bar{t}}$, reflecting the parton distribution functions of the incoming
protons and antiprotons.

The resolution on the \tt invariant mass is limited by the  
uncertainty on the jet energy and by the unmeasured longitudinal
momentum of the neutrino.
The approach adopted here is to use a matrix element technique \cite{DLM} to
reconstruct \Mtt for an event.
We integrate the matrix-element for standard model 
\tt production over unmeasured quantities, convoluted with detector 
resolution functions for the jet energies.
The matrix element implemented for the event reconstruction is the
standard model leading order \tt production and decay,
$q\bar{q} \, (gg) \rightarrow t\bar{t}\rightarrow W^{+}bW^{-}\bar{
b}\rightarrow  q \bar{q}' b \ell \nu_{\ell} \bar{b}$, and
it is used to derive the value of \Mtt in the same
way for events in simulated samples and in data.
%

The \emph{a priori} probability density for producing a \tt
parton-level final state, $\pi_{\rm{part}}(\{\vec{p}\})$,
relative to other \tt final states, is the normalized differential cross section
\begin{multline*}
\pi_{\rm{part}}(\{\vec{p}\}) \prod_i d^{3}\vec{p_{i}}=\\
=\frac{1}{\sigma}\int_0^1\int_0^1 dz_{a} dz_{b} f_{k}(z_{a})f_{l}(z_{b})d\sigma_{kl}(\{\vec{p}\}, z_{a}\vec{P},z_{b}\vec{\bar{P}}),
\end{multline*}
where $f_k$ ($f_l$) are the proton (antiproton) parton distribution functions,
$d\sigma_{kl}$ the \tt differential cross section, $\vec{P}~(\vec{\bar{P}})$
the proton (antiproton) momentum and $\{\vec{p}\}$ the set of six final 
state three-momenta.
Indices \textit {k, l} cover the parton types in the proton and
antiproton respectively, and a sum over both indices is implied.
The
\emph{a priori} probability density for the parton level final state $\{\vec{p}\}$
and the measured jet quantities $\{\vec j\}$ is given by the product
\begin{multline*}
\pi_{\rm{part,jets}}(\{\vec{p}\}, \{\vec{j}\}) =\\
=\pi_{\rm{part}}(\{\vec{p}\})\cdot T(\vec{j}_1|\vec{p}_1) \cdot T(\vec{j}_2|\vec{p}_2)
\cdot T(\vec{j}_3|\vec{p}_3) \cdot T(\vec{j}_4|\vec{p}_4),
\end{multline*}
where
$T(\vec{j}|\vec{p})$ denotes the parton to jet transfer
function, i.e. the probability that a parton of momentum $\vec{p}$ is
measured as a jet of momentum $\vec{j}$.
Transfer functions are parametrized in $|\eta|$ and momentum 
of the jet, and are derived
from Monte Carlo simulated ({\sc pythia}) \cite{pythia} events. 
Different sets of transfer functions are
adopted for $b$-quarks and lighter quarks.
No transfer function is used for the charged lepton given its negligible
momentum uncertainty.
Also, the \met\, measurement is not used to constrain the neutrino
transverse momentum.

From $\pi_{\rm{part,jets}}(\{\vec{p}\}, \{\vec{j}\})$  we build $P(\{\vec{p}\}|
\{\vec{j}\})$, the probability density for the parton momenta 
$\{\vec{p}\}$, given 
the observed quantities  $\{\vec{j}\}$. 
From this distribution
we derive probability distributions for the new variable, $M_{t\bar{t}}$,
which is a function
of the parton level quantities $\{\vec{p}\}$,
by calculating
$P_f(x| \{\vec{j}\}) = \int{P(\{\vec{p}\}|
\{\vec{j}\}) \cdot \delta(x - M_{t\bar{t}}(\{\vec{p}\}))}~d\{\vec{p}\}$.
Having an event probability distribution for \Mtt we choose the
mean as the reconstructed value for that event, since Monte
Carlo simulations showed it to be the
best single value $M_{t\bar t}$ estimator. However, given 
that we do not know which jet
matches which parton we also sum over all possible permutations before
extracting the mean. If the event contains jets identified as originating from 
$b$-quarks we sum only over permutations with $b$-tagged jets assigned to 
$b$-quark partons.

To produce the \Mtt templates
we apply the reconstruction algorithm to each
signal and background sample.
The {\sc pythia} \cite{pythia} generator was used to simulate both the
$Z'$ vector resonant production ($M_t = 175 ~{\rm{GeV}}/c^2$)
and \tt events.
{\sc alpgen} \cite{alpgen} was used for the simulation of the $W$
boson plus parton production, with {\sc herwig} \cite{herwig}
used to model parton showers. 
All generated events are passed through the CDF detector simulation.
The multi-jet background template was obtained using CDF II data.

\begin{figure}[htb]
\vspace{-0.3cm}
\centerline{
\resizebox{8cm}{!}{\includegraphics{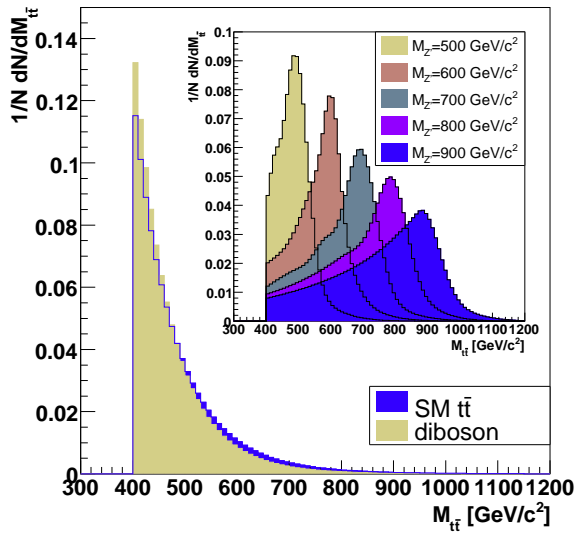}}
}
\caption{Reconstructed \Mtt distributions for two simulated standard model
backgrounds. We plot the samples with the ``hardest'' ($t\bar{t}$) and the
``softest'' (diboson) spectrum (normalized to unity); all the others lie in
between. The inset shows the reconstructed \Mtt distributions for five signal
samples.} 
\label{fig:BandS}
\end{figure}

To allow direct comparison with the
Tevatron Run I results \cite{Run1X0} we choose a resonance width
$\Gamma_{Z'} = 0.012 ~M_{Z'}$; however, the measured signal cross section
upper limits are insensitive to width values up to $\Gamma_{Z'} \approx 0.05
~M_{Z'}$.
The algorithm shows an intrinsic resolution on \Mtt of about 25
${\rm{GeV}}/c^2$  when
applied to standard model \tt events, in case of correct partons to
jet association.
When applied to resonance samples the mass is still correctly reconstructed
in the case of proper  parton-jet assignment; however, incorrect parton-jet
assignment 
leads to a low tail in the reconstructed mass distribution as explained in
Fig. \ref{fig:BandS}.
The spectra plotted are restricted to the search region 
(\Mtt$>400~{\rm{GeV}}/c^2$) in which the standard model sources fall off
exponentially in $M_{t\bar{t}}$.
%
%
%
%

To derive the posterior probability for
$\sigma(p\bar p \rightarrow Z')\cdot BR(Z' \rightarrow t\bar t )$ 
(we will refer to it simply as ``signal cross section'')
given the
observed \Mtt spectrum, we build the likelihood
$
\mathcal{L} (\vec n |\sigma,{\vec \nu}) =
\prod_{i\in\{bins\}} \, e^{-\mu_i} {\mu_i^{n_i}}/{n_i!} 
$.
This is the prior probability of observing $\vec{n}$, where $n_i$ 
is the number of
observed events in \Mtt mass bin \textit{i}, 
$\mu_i=\sigma_s A_s T_s{_i} + \sum_j N_{j} T_j{_i}$ 
is the number of expected
events in the same bin, $\sigma_s$ is the assumed signal cross
section, 
$A_s$ is the acceptance, $N_{j}$ is the contribution 
from the $j^{th}$
background and $T_s{_i}$, $T_j{_i}$ is the content of the 
$i^{th}$ bin of the normalized
signal and background templates, respectively. 
The contributions from standard model \tt and dibosons are
weighted as in Table \ref{tab:Backgrounds}, while
the contributions due to \emph{W}+jets and multi-jets are rescaled
to account for the presence of signal and still match the total number of 
events observed
($N_{W+\rm{jets}}$+$N_{\rm{multi-jet}}$=$N_{\rm{data}}-N_{\rm{signal}}-N_{t\bar t}-N_{\rm{diboson}}$).
If one uses a flat prior distribution for the signal cross section and
integrates over the nuisance
parameters $\vec \nu$ (background contributions and signal acceptance)
then Bayes' theorem gives the posterior probability density $P(\sigma|\vec n)$.

The posterior probability density is used to define upper 
limits at any given confidence
level (CL), together with the most likely value which is regarded as the
measured cross section. 
This procedure is repeated for 10 resonance masses from 450 ${\rm{GeV}}/c^2$
to 900 ${\rm{GeV}}/c^2$ and 95\% CL upper limits are established.
%
%

As mentioned previously, uncertainties on template
weighting are incorporated in the prior
probability by means of the prior densities of the nuisance parameters. 
In this way the marginalized posterior probability density for the signal cross section
includes the acceptance and cross section uncertainties.
%
%
%
%
However, the relative contribution of multi-jet and \emph{W}+jet events 
has been kept fixed at 10\%. To evaluate the impact of the uncertainty 
on this ratio, the multi-jet component has been either set to zero or 
increased by a factor of three, 
in each case yielding negligible change on the cross section
posterior distribution. This is because after the event selection 
the \emph{W}+jet and multi-jet \Mtt spectra are very similar above $400
~{\rm{GeV}}/c^2$, as shown in Fig. \ref{fig:BandS}.
We assume a 10\% acceptance uncertainty for all the resonant \tt signals. 

Other sources of systematic uncertainties
can affect both the
acceptances and the templates. 
These ``shape systematics'' are due to the imperfect knowledge
in the modeling of
(1) jet energy scale, 
(2) ISR and FSR, 
(3) the $Q^2$ scale for \emph{W}+jets production, and
(4) parton distribution functions (PDF).
To account for these uncertainties we convolute the cross section
posterior probability with a Gaussian, whose width is
estimated from a new set of templates and acceptances corresponding to 
a $\pm 1\sigma$ change for each systematic.
The result of this convolution smears the cross
section posterior, pushing the upper limits to higher cross section values.
For a resonance of mass $M_{Z'} = 700~{\rm{GeV}}/c^{2}$ the uncertainties on 
the jet energy scale, initial and final state radiation, $W-Q^2$
scale translate into a relative increase of the expected upper limits of
5\%, 5\%, 2\% and 2\% respectively, for an overall impact of about 10\%.
The uncertainty due to the choice of the PDFs turned out to be negligible. 

%

\begin{figure}[htb]
\vspace{-0.3cm}
\centerline{
\resizebox{8cm}{!}{\includegraphics{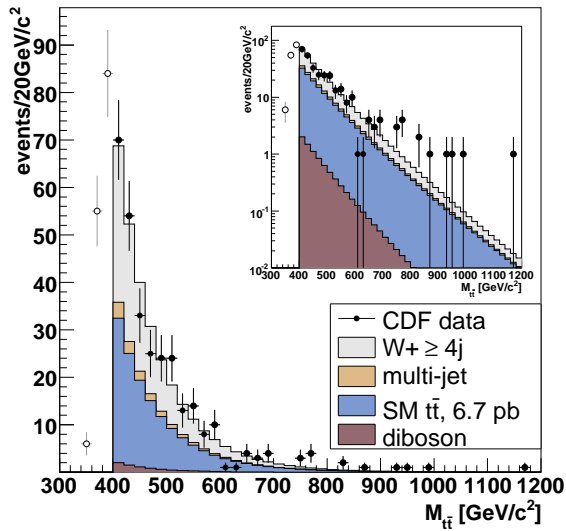}}
}
\caption{The reconstructed \Mtt spectrum (data) and the standard model prediction in the search region above 400 ${\rm{GeV}}/c^2$. 
The inset shows the same distributions in logarithmic scale.
}
\label{fig:Data}
\end{figure}

In the data sample used for this analysis, 450 events passed our 
event selection requirements, and 302 of them are found in the search region 
$M_{t\bar{t}} > 400~{\rm{GeV}}/c^2$ .
The \Mtt data spectrum is shown in Fig. \ref{fig:Data}, together with the
background expectations (above 400 ${\rm{GeV}}/c^2$) based on Monte Carlo studies.
In order to establish an \emph{a priori} sensitivity of the reconstruction
algorithm we generated 1000 simulated experiments in the
null hypothesis and extracted, for each mass point, the 95\% CL expected 
upper limit, defined as the median of the upper limits distribution. 
We also calculated the central 1$\sigma$ and 2$\sigma$ frequentist coverage
bands.

\begin{figure}[htb]
\vspace{-0.3cm}
\centerline{
\resizebox{8.2cm}{!}{\includegraphics{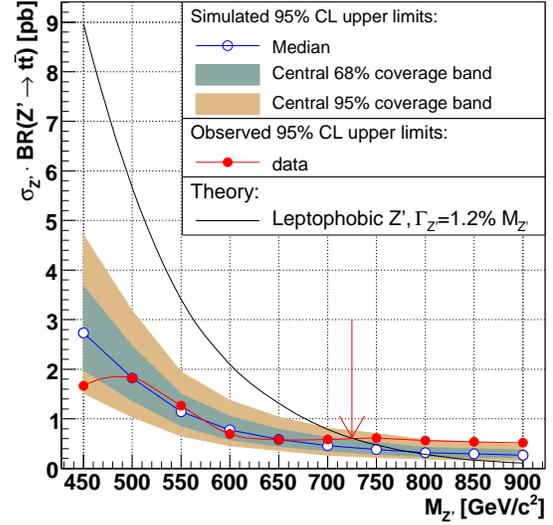}}
}
\caption{Predicted and experimental 95\% CL upper limits using data corresponding to $680 ~\rm{pb^{-1}}$ of integrated luminosity, together with the leptophobic topcolor $Z'$ cross section prediction. Dark and light areas define the $1 \sigma$ and $2 \sigma$ ranges for the expected limits. The arrow marks the mass upper limit.}
\label{fig:Brazil}
\end{figure}

In the data, the posterior probability distributions for the  
signal cross section show no evidence of resonant \tt production.
The predicted and observed upper limits at 95\% CL are shown in 
Fig. \ref{fig:Brazil},
together with the theoretical prediction for the cross section as a function of
mass in the leptophobic $Z'$ model (with $\Gamma_{Z'}=0.012 ~M_{Z'}$)
\cite{Harris}.
Based on this model we exclude a leptophobic
topcolor resonance candidate with a mass of 725 ${\rm{GeV}}/c^2$ or less.


In conclusion, we performed a search for a narrow heavy resonance decaying into \tt
in the lepton+jets channel using $680 ~\rm{pb^{-1}}$ of CDF Run 2 data.
We set upper limits on the production cross section times branching ratio
at the 95\% CL.  For one leptophobic topcolor production mechanism we exclude
masses up to 725 ${\rm{GeV}}/c^2$, extending significantly the Run I
limit of 560 ${\rm{GeV}}/c^2$ \cite{Run1X0}.

We thank the Fermilab staff and the technical staffs of the participating
institutions for their vital contributions. This work was supported by the
U.S. Department of Energy and National Science Foundation; the Italian Istituto
Nazionale di Fisica Nucleare; the Ministry of Education, Culture, Sports,
Science and Technology of Japan; the Natural Sciences and Engineering Research
Council of Canada; the National Science Council of the Republic of China; the
Swiss National Science Foundation; the A.P. Sloan Foundation; the
Bundesministerium f\"ur Bildung und Forschung, Germany; the Korean Science and
Engineering Foundation and the Korean Research Foundation; the Science and
Technology Facilities Council and the Royal Society, UK; the Institut National
de Physique Nucleaire et Physique des Particules/CNRS; the Russian Foundation
for Basic Research; the Comisi\'on Interministerial de Ciencia y
Tecnolog\'{\i}a, Spain; the European Community's Human Potential Programme; the
Slovak R\&D Agency; and the Academy of Finland. 

\begin{thebibliography}{99}
\bibitem{XYZ} 
CDF uses a $(z,\,\phi,\,\theta)$ coordinate system with 
the $z$-axis in the direction of the proton beam; $\phi$ and
$\theta$ are the azimuthal and polar angle respectively. The
pseudorapidity is defined as $\eta= -\rm{ln(tan}\frac{\theta}{2})$, 
and the transverse momentum and energy as $p_T = p \sin \theta$ and
$E_T = E \sin \theta$ respectively.
Missing transverse energy is defined as
$\not\!\! \vec{E}_T = - \sum_{i} E_T^i {\bf \hat{n}}_i$, where
${\bf \hat{n}}_i$ is a unit vector in the transverse plane that points
from the beam-line to the $i^{th}$ calorimeter tower.
\end{thebibliography}
\end{document}